\documentclass{article}

\usepackage[preprint]{neurips_2024}

\usepackage[utf8]{inputenc} % allow utf-8 input
\usepackage[T1]{fontenc}    % use 8-bit T1 fonts
\usepackage{hyperref}       % hyperlinks
\usepackage{url}            % simple URL typesetting
\usepackage{booktabs}       % professional-quality tables
\usepackage{amsfonts}       % blackboard math symbols
\usepackage{nicefrac}       % compact symbols for 1/2, etc.
\usepackage{microtype}      % microtypography
\usepackage{xcolor}         % colors

\usepackage{graphicx}
\usepackage{amsmath}
\usepackage{amsthm}
\usepackage{algorithm}
\usepackage{algpseudocode}
\usepackage{multirow}

\newtheorem{lemma}{Lemma}
\newtheorem{remark}{Remark}

\title{Adaptive Bayesian Multivariate Spline Knot Inference with Prior Specifications on Model Complexity}

\author{%
  Junhui He \\
  Department of Mathematical Sciences\\
  Tsinghua University\\
  Beijing, China \\
  \texttt{hejh22@mails.tsinghua.edu.cn} \\
  \And
  Ying Yang\\
  Department of Mathematical Sciences\\
  Tsinghua University\\
  Beijing, China \\
  \texttt{yangying@tsinghua.edu.cn} \\
  \And
  Jian Kang \\
  Department of Biostatistics\\
  University of Michigan, Ann Arbor\\
  Michigan, United States\\
  \texttt{jiankang@umich.edu}\\
}

\begin{document}

\maketitle

%% mainbody

\begin{abstract}
%% Text of the abstract
In multivariate spline regression, the number and locations of knots influence the performance and interpretability significantly. However, due to non-differentiability and varying dimensions, there is no desirable frequentist method to make inference on knots. In this article, we propose a fully Bayesian approach for knot inference in multivariate spline regression. The existing Bayesian method often uses BIC to calculate the posterior, but BIC is too liberal and it will heavily overestimate the knot number when the candidate model space is large. We specify a new prior on the knot number to take into account the complexity of the model space and derive an analytic formula in the normal model. In the non-normal cases, we utilize the extended Bayesian information criterion to approximate the posterior density. The samples are simulated in the space with differing dimensions via reversible jump Markov chain Monte Carlo. We apply the proposed method in knot inference and manifold denoising. Experiments demonstrate the splendid capability of the algorithm, especially in function fitting with jumping discontinuity.
\end{abstract}

\section{Introduction}

Spline regression \citep{wahba1990spline, schumaker2007spline, gu2013smoothing} is a nonparametric method for modelling the complex dependencies between features, widely used in many fields including machine learning, econometrics and biomedicine. It is an ideal alternative to linear regression in the nonlinear data analysis. However, given the number and location of knots, the spline space is actually a linear space with a spline basis. Thus, the spline regression degenerates to a simple linear regression with respect to the basis, posing a strict limitation on its representative capacity. The ordinary solution is to assign sufficiently many knots and locate them uniformly, leading to a trade-off between the complexity and the flexibility of splines; see smoothing splines and thin plate splines \citep{10.1111/1467-9868.00374}. Furthermore, the splines with fixed knots usually assume each knot is used only once. This means the spline function has continuous derivatives up to its degree minus one. Consequently, it is inappropriate for the distinct-knot spline to fit curves with jumping discontinuity. %illustrated in Figure \ref{fig:spline}.

\iffalse
\begin{figure}
    \centering
    \includegraphics[width=0.5\textwidth]{Experiments/plots/spline_regression.pdf}
    \caption{A cubic spline with the jumping discontinuity at $0.4$.}
    \label{fig:spline}
\end{figure}
\fi

Rather than equally-spaced knots, we develop a Bayesian approach for the automatic selection of the spline knots. 
%In the multiple regression case, we utilize the tensor product spline \citep{dierckx1995curve} to make predictions.
% Rather than equally-spaced or quantile-based knots, we propose an extended Bayesian adaptive knot spline regression method (EBARS) to estimate the proper number and location of knots. 
The basic principle is modelling the intricate relationships with a few knots and the optimal location. The adaptive knot method reduces the complexity of splines when preserving the flexibility. We can also model the sharp changes by placing several knots in the almost coincident position. Moreover, the knot estimation provides a mechanism for easy interpretation. The number of knots detects how many transitions occur and the location indicates where change points lie. Thus, the quantities are of independent interest in many applications. For example, by examining the approval ratings of successful politicians, the president can determine the optimal time to switch from the job of governing to the job of running for reelection \citep{marsh2001spline}. See \citet{aminikhanghahi2017survey, truong2020selective} for a review about the change point detection in time series.

A considerable amount of work has been done for the knot selection. Some researchers rely on specialist knowledge or exploratory data analysis to specify the knots heuristically. But the ad hoc manner is rough since the exact knot information is unknown. \citet{10.2307/2346413} chose the knots via a grid-search method, whose search time grows exponentially with the increasing sample size. \citet{muggeo2003estimating} proposed a segmented regression method to estimate the break-point location based on a linearization technique and developed an R package \citep{muggeo2008segmented}. But this method is limited to linear splines and the number of knots is required to be known. For likelihood-based inference, the primary difficulty is that the likelihood is not differentiable with respect to the knots. Given the number of knots, \citet{doi:10.1080/01621459.2015.1073154} utilized local quadratic smoothing to approximate linear spline models and \citet{doi:10.1080/01621459.2021.1947307} proposed modified derivatives of the likelihood at the knots. However, the knot number cannot be estimated in the two methods.

Bayesian approaches were also proposed in the literature. \citet{https://doi.org/10.1111/1467-9868.00128} proposed an automatic Bayesian curve fitting method and extended it to multivariate splines in \citet{Denison1998MARS}. However, they did not derive the correct dimensional penalty factor and the method involved hyperparameter determination. \citet{10.1093/biomet/88.4.1055} developed a Bayesian adaptive regression spline method for the knot estimation. They employed the Bayesian information criterion (BIC) to calculate the marginal likelihood and simulated samples via reversible jump Markov chain Monte Carlo (RJMCMC) of \citet{10.1093/biomet/82.4.711}. However, the prior of the knot number doesn't take into account the complexity of the model space when the parametric dimension changes, and BIC seems to be too liberal as evaluation criteria when the number of all possible knots is large. Besides, this method is restricted to univariate splines. \citet{fearnhead2006exact} proposed a recursive approach to calculate the density and performed direct simulation from the posterior distribution. But the computational complexity is quadratic in the sample size, making it a slow method. Refer to \citet{chen2011comparison} for a detailed comparison about the Bayesian knot estimation. %Except for \citet{10.1093/biomet/88.4.1055}, other methods mentioned fall into a predicament when the jumping discontinuity occurs.

In this article, we will propose an extended Bayesian adaptive regression spline (EBARS) method for estimating the number and location of multivariate spline knots simultaneously. The tensor product spline \citep{dierckx1995curve} is used to make predictions in the multiple regression case. We consider the complexity of the model space and thereby assign a specific prior on the number of knots to adjust the effect. We specify a unit information prior \citep{e447aec4-501c-3132-a415-1b260054b4a8} for the spline coefficients. An analytic expression of the evidence is derived in the normal noise model. For the non-normal cases, we utilize the extended Bayesian information criterion (EBIC) for the approximation. Similar to \citet{10.1093/biomet/88.4.1055}, the posterior simulation is performed via RJMCMC. We compare the performance with several methods in knot inference through extensive simulations. Experiments demonstrate the excellent performance of our method in scenarios with single or multiple knots, with or without jumping discontinuity. Furthermore, we develop a manifold estimation technique as an application of the proposed method. An R package is available in the GitHub repository \url{https://github.com/junhuihe2000/EBARS}.

The paper is organized as follows: Section 2 presents the detailed Bayesian spline knot selection method; Section 3 describes the sampling procedure of RJMCMC; Section 4 evaluates the proposed algorithm in knot inference and manifold denoising through numerical experiments; Section 5 concludes and discusses the future research.
\section{Extended Bayesian adaptive regression spline}

We are considering the tensor product spline for modeling the dependencies between a single response variable and multiple predictor variables. Generally, the number and locations of knots determine the smoothness of splines and affect the performance substantially. Our objective is to select optimal nodes by a Bayesian approach.

Assume labeled observations $\{(x_i,y_i)\}_{i=1}^m\subset [0,1]^d\times \mathbb{R}$ are independent and identically distributed, such that for $i=1,\ldots,m$,
\begin{equation}
\label{eq:normal_regression}
    y_i = f(x_i) + \epsilon_i,\quad \epsilon_i\sim N(0,\sigma^2),
\end{equation}
where $f:[0,1]^d\rightarrow \mathbb{R}$ is an unknown function, and $\sigma^2>0$ is the noise variance.

\subsection{Tensor product spline regression model}

The article utilizes tensor product splines to model the real multivariate function $f$. As the name implies, the tensor product spline space is the tensor product of $d$ univariate spline spaces corresponding to each component of $x$. The univariate spline space in $[0,1]$ of degree $p_i$ and non-decreasing knot sequence $\{\xi_{ij}\}_{j=0}^{k_i+1}$ allowing duplicates $(\xi_{i0}=0,\xi_{i(k_i+1)}=1)$ consists of the following univariate functions: (1) Be a polynomial of at most degree $p_i$ on each interval $(\xi_{ij},\xi_{i(j+1)})$; (2) Belong to $C^{p_i-1}([0,1])$ except for the coincident knots, and if $\xi_{i(j-1)}<\xi_{ij}=\ldots=\xi_{i(j+l)}<\xi_{i(j+l+1)}$, the function has continuous derivatives up to the order $p_i-1-l$ at that point $(l\leq p_i)$. Specifically, when $l=p_i$, the spline function may discontinue at the coincident knot.
%illustrated in Figure \ref{fig:spline}. 
%We refer readers to \citet{dierckx1995curve} for a comprehensive review of the splines with potential duplicate knots.

\begin{remark}
    The usual spline-based methods employ equidistant knots or quantile-based knots, posing the distinct knot assumption implicitly. This results in continuous splines. When the true function exists jumping discontinuity, the distinct knot spline will fail to make an accurate estimation. However, the spline with automatic knot selection can circumvent the problem by optimal placement.
\end{remark}

% As the boundary knots are given by 0 and 1, the unknown knots are $\{\xi_{ij}\}_{j=1}^{k_i}$. 
Let $p=\{p_i\}_{i=1}^d$, $k=\{k_i\}_{i=1}^d$, and $\xi=\{\xi_i\}_{i=1}^d$ with $\xi_i=\{\xi_{ij}\}_{j=1}^{k_i}$. Denote the tensor product spline space as $\mathcal{S}_{p,k,\xi}$ and the univariate spline space as $\mathcal{S}_{p_i,k_i,\xi_i}$ for $i=1,\ldots,d$. Then $\mathcal{S}_{p,k,\xi}=\bigotimes_{i=1}^d \mathcal{S}_{p_i,k_i,\xi_i}$, where $\mathcal{S}_{p_i,k_i,\xi_i}$ is a linear space with the dimension of $k_i+p_i+1$. Let $\{b_{ij}\}_{j=1}^{k_i+p_i+1}$ be a basis of $\mathcal{S}_{p_i,k_i,\xi_i}$, such as B-splines. Consequently, the dimension of $\mathcal{S}_{p,k,\xi}$ is $\Pi_{i=1}^d (k_i+p_i+1)$, denoted as $\nu$. And $b=\bigotimes_{i=1}^d \{b_{ij}\}_{j=1}^{k_i+p_i+1}$ is a basis of $\mathcal{S}_{p,k,\xi}$. For simplicity, rewrite $b$ as $\{b_i\}_{i=1}^\nu$. Thus, any $f\in \mathcal{S}_{p,k,\xi}$ can be represented by $\sum_{i=1}^\nu \beta_i b_i$. Let $\beta=(\beta_1,\ldots,\beta_\nu)^\top$. Define the design matrix $Z$ by $Z_{ij}=b_j(x_i)$ for $i=1,\ldots,m$ and $j=1,\ldots,\nu$. Therefore, \eqref{eq:normal_regression} can be reformulated as
\begin{equation}
    \label{eq:spline_regression}
    y = Z\beta + \epsilon,\quad \epsilon\sim N_m(0,\sigma^2 I_m),
\end{equation}
where $y,\epsilon \in \mathbb{R}^m$, $\beta\in \mathbb{R}^\nu$, $Z\in \mathbb{R}^{m\times \nu}$. Notably, \eqref{eq:spline_regression} is an ordinary linear regression model.
%Thereby we can use those consequences of linear regression to make inferences.
\subsection{Prior \& posterior}

% When $p,k,\xi$ are fixed, \eqref{eq:spline_regression} implies the linear dependencies. This restricts the modeling capacity. 
Experiments have demonstrated that changes on $p$ have little impact on the performance; see \citet{perperoglou2019review}. Cubic splines $(p=3)$ and linear splines $(p=1)$ are the most common alternatives in the research. We will explore the two splines in the simulations. Given the degree, the number and position of knots determine basis functions of the spline space, affecting the non-linearity in the model. 
%However, the non-differentiability of the likelihood on $k$ and $\xi$ causes a primary challenge in the estimation, prohibiting the gradient-based technique for optimization. Frequentist researchers utilize the modified derivatives or construct the smooth approximations at the non-differentiable points to solve the issue partially; see \citet{doi:10.1080/01621459.2015.1073154, doi:10.1080/01621459.2021.1947307}. They cannot handle the estimation of $k$ and the methods are complex. In contrast, Bayesian approaches provide a universal framework to resolve the varying dimension problem of parameters by reversible jump Markov chain Monte Carlo (RJMCMC) of \citet{10.1093/biomet/82.4.711}. 
%In this section, we develop a Bayesian method to estimate $k,\xi$.

We specify the priors of $k,\xi,\beta,\sigma$ in \eqref{eq:spline_regression} as follows. Firstly, we initialize enormous candidate knots. Owing to the specific structure of tensor product splines, the candidate knots can be chosen in each component separately. Let $n_i$ nodes $\eta_i\subset [0,1]$ be the $i$-th component. Then the overall candidate knots are given by $\eta=\bigotimes_{i=1}^d \eta_i$. To simplify the analysis, assume that the knots in $\eta_i$ are distinct. Nevertheless, when $n_i$ is sufficiently large and $\eta_i$ is sufficiently dense in $[0,1]$, we can still find a spline model to reflect jumping discontinuity well. %This is different from the smoothing spline case where excessive knots lead to overfitting. 
For $i=1,\ldots,d$, let $\mathcal{M}_{k_i}$ be the model space containing all the possible location combinations with $k_i$ knots in $[0,1]$. Then the size of $\mathcal{M}_{k_i}$ is $\tau(\mathcal{M}_{k_i})=\binom{n_i}{k_i}$. Since $k_i<<n_i$, $\tau(\mathcal{M}_{k_i})$ is increasing as $k_i$ increases. Let $\mathcal{M}_k$ be $\bigotimes_{i=1}^d \mathcal{M}_{k_i}$. Then $\tau(\mathcal{M}_k)=\Pi_{i=1}^d \binom{n_i}{k_i}$. The priors of $k,\xi$ are specified as 
\begin{equation}
    \label{eq:prior}
    \pi(k)\propto \tau(\mathcal{M}_k)^{1-\gamma},~\pi(\xi|k)={1}/{\tau(\mathcal{M}_k)},\quad 0\leq \gamma \leq 1.
\end{equation}
Consequently, $\pi(k,\xi)\propto \tau(\mathcal{M}_k)^{-\gamma}$. Given $k$, the prior probabilities of different position combinations are equal. The parameter $\gamma$ adjusts the growth rate of $\pi(k)$. It is equivalent to imposing a penalty on $\pi(k)$. As $\gamma$ increases from 0 to 1, the growth rate becomes milder gradually. Particularly, when $\gamma=0$, we have $\pi(k)\propto \tau(\mathcal{M}_k)$ and $\pi(k,\xi)\propto 1$, meaning the same prior probability for knots. Sometimes this $\gamma$ causes that excessive knots are selected.

Suppose $k_1,\ldots,k_d<<m$, then $Z^\top Z$ is of full rank. In terms of $\beta,\sigma$, we assume
\begin{equation}
    \label{eq:prior_beta}
    \beta|Z,\sigma \sim N_\nu\left(0,m\sigma^2(Z^\top Z)^{-1}\right),~\pi(\sigma)={1}/{\sigma},\quad \sigma>0.
\end{equation}
The prior of $\beta$ is the so-called unit information prior. According to the linear regression theory, the least squares estimator of $\beta$ is $\hat{\beta}=(Z^\top Z)^{-1}Z^\top y$. Then the precision matrix (inverse covariance) is given by $(Z^\top Z)/\sigma^2$. The unit precision matrix is defined as $(Z^\top Z)/m\sigma^2$, implying the prior of $\beta$ in \eqref{eq:prior_beta}. %Please refer to \citet{e447aec4-501c-3132-a415-1b260054b4a8} for detailed discussions about the unit information prior. 
The prior of $\sigma$ is an improper prior as $\int_0^\infty 1/\sigma d\sigma=\infty$. % In fact, $\pi(\sigma)$ is not unique but $1/\sigma$ is the simplest case.

According to the Bayesian formula, the posterior density of $k,\xi,\beta,\sigma$ is
\begin{equation}
    \label{eq:jointposterior}
    p(k,\xi,\beta,\sigma|y)=p(\beta,\sigma|k,\xi,y)p(k,\xi|y),
\end{equation}
where $p(k,\xi|y)\propto p(y|k,\xi)\pi(k,\xi)$. Let $a_{k,\xi}=y^\top (I_m-\frac{m}{m+1}Z(Z^\top Z)^{-1}Z^\top)y$.

\iffalse
According to the Bayesian formula, the posterior density of $k,\xi$ is $p(k,\xi|y)\propto p(y|k,\xi)\pi(k,\xi)$. And $p(y|k,\xi)=p(y|Z)$ is the marginal likelihood given by
\begin{equation}
    \label{eq:likelihood}
    p(y|Z) = \int_{(0,\infty)} \int_{\mathbb{R}^\nu} p(y|Z,\beta,\sigma) \pi(\beta|Z,\sigma)\pi(\sigma) d\beta d\sigma.
\end{equation}
Let $a_{k,\xi}=y^\top (I_m-\frac{m}{m+1}Z(Z^\top Z)^{-1}Z^\top)y$.
\fi

\begin{lemma}
\label{lm:post}
    With the above priors of $k,\xi,\beta,\sigma$ in \eqref{eq:spline_regression}, we have
    \begin{equation}
        \label{eq:posterior}
        p(y|k,\xi)\propto (m+1)^{-\nu/2}a_{k,\xi}^{-m/2},~ p(k,\xi|y)\propto (m+1)^{-\nu/2}a_{k,\xi}^{-m/2} \tau(\mathcal{M}_k)^{-\gamma}.
    \end{equation}
\end{lemma}

The posterior $p(k,\xi|y)$ consists of three main components. The first term $(m+1)^{-\nu/2}$ serves as the dimensional penalty. It balances the number of parameters and the bias of model fitting. The second term $a_{k,\xi}^{-m/2}$ represents the effect of the likelihood. When $m$ is sufficiently large, $a_{k,\xi}$ is approximately equal to the residual sum of squares. The third term $\tau(\mathcal{M}_k)^{-\gamma}$ corresponds to the priors of $k,\xi$. It takes the complexity of $\mathcal{M}_k$ into consideration. %The hyper-parameter $\gamma\in [0,1]$ adjusts the effect of $\tau(\mathcal{M}_k)$.

Subsequently, we can simulate samples of $\beta,\sigma$ given $k,\xi$ from the conditional posterior density in \eqref{eq:jointposterior} via a Gibbs sampler, contributing to a fully Bayesian model.

%According to Lemma \ref{lm:post}, $k,\xi$ are estimated by the posterior distribution in \eqref{eq:posterior}. Subsequently, we can construct the maximum likelihood estimation or least squares estimation for $\beta,\sigma$. The posterior expectations of $\beta,\sigma$ are also practicable, contributing to a fully Bayesian model. We adopt the maximum likelihood estimation in the paper.

%\citet{10.1093/biomet/88.4.1055} proposed a Bayesian technique for the adaptive knot spline regression as well. However, their method is limited to the univariate splines, whereas our EBARS are extended to the multi-dimensional cases by tensor product splines. Besides, the priors of $k$ are different. They don't consider the effect of $\tau(\mathcal{M}_k)$ when the parametric dimension varies. This poses an incorrect estimation when the true $k$ is huge, illustrated in the simulation examples.

\subsection{Extended Bayesian information criterion}

The Gaussian prior of $\beta$ is a conjugate prior in \eqref{eq:spline_regression}, yielding a closed expression of $p(y|k,\xi)$. Generally, $p(y|k,\xi)$ is analytically intractable except for the normal regression model. %Then we have to calculate the marginal likelihood numerically. 
For those non-conjugate cases, we utilize the extended Bayesian information criterion (EBIC) of \citet{10.1093/biomet/asn034} to approximate the posterior density. %Informally, the EBIC takes into account the dimension of unknown parameters and the complexity of the model space, extremely useful for variable selection in problems with a huge number of candidate covariates. %\citep{NIPS2010_072b030b, bb6df906-631b-39ae-98e5-1c3a3cd1049c, Luo2015}.
In the spline knot estimation, the cardinality of all candidate knots (\emph{i.e.}, $n$) can be very large but the number of the true knots (\emph{i.e.}, $k$) is small compared to the sample size (\emph{i.e.}, $m$). Thus, EBIC will be extremely useful for model selection as the small-$m$-large-$n$ assumption holds and the Laplace approximation is valid \citep{NIPS2010_072b030b, bb6df906-631b-39ae-98e5-1c3a3cd1049c, Luo2015}.
% give an excellent estimation of the posterior density.

According to the definition, the EBIC of $k,\xi$ is
\begin{equation}
    \label{eq:ebic}
    \text{BIC}_{\gamma}(k,\xi) = -2\log L(\hat{\beta},\hat{\sigma}|y,k,\xi) + (\nu+1) \log m + 2\gamma \log \tau(\mathcal{M}_k),\quad 0\leq \gamma \leq 1,
\end{equation}
where $\hat{\beta},\hat{\sigma}$ are the maximum likelihood estimators of $\beta,\sigma$ given $k,\xi$. Especially in \eqref{eq:spline_regression}, $\hat{\beta}=(Z^\top Z)^{-1}Z^\top y$ and $\hat{\sigma}^2 = y^\top (I_m-Z(Z^\top Z)^{-1}Z^\top)y/m$. %And $\gamma\in [0,1]$ is a hyper-parameter of the penalty term involved with the complexity of $\mathcal{M}_k$. 
As $\gamma=0$, \eqref{eq:ebic} is the ordinary BIC. Since $n_1,\ldots,n_d>>m$, the EBIC with $\gamma>0$ is preferable to the ordinary BIC in knot estimation. From the Laplace approximation, $p(k,\xi|y)\approx \exp\{-\text{BIC}_\gamma(k,\xi)/2\}$, denoted by $\hat{p}(k,\xi|y)$.
%Furthermore, $\text{BIC}_\gamma(k,\xi)$ is the Laplace approximation of $-2\log p(k,\xi|y)$. Thus, $p(k,\xi|y)\approx \exp\{-\text{BIC}_\gamma(k,\xi)/2\}$, denoted by $\hat{p}(k,\xi|y)$.

\begin{lemma}
    In multivariate spline model \eqref{eq:spline_regression}, the EBIC approximation of the posterior density is
    \begin{equation}
        \label{eq:approx}
        \hat{p}(k,\xi|y)\propto m^{-(\nu+1)/2} (\hat{\sigma}^2)^{-m/2} \tau(\mathcal{M}_k)^{-\gamma}.
    \end{equation}
\end{lemma}
Suppose $k',\xi'$ are another group of knots. Comparing \eqref{eq:posterior} and \eqref{eq:approx}, we can find $p(k,\xi|y)/p(k',\xi'|y)\approx \hat{p}(k,\xi|y)/\hat{p}(k',\xi'|y)$ when $m$ is sufficiently large. Since the sampling procedure of MCMC is determined by the posterior density ratio, the EBIC contributes to a consistent estimation.
\section{Reversible jump Markov chain Monte Carlo}

The reversible jump approach \citep{10.1093/biomet/82.4.711} is an extension of the standard Metropolis-Hastings algorithm, allowing the trans-dimensional movement. These algorithms are widely used in the (Bayesian) model determination problems where the dimension of parameters is unknown \citep{doi:10.1080/01621459.2018.1423984, https://doi.org/10.1002/sim.8511}. 

In this section we use RJMCMC to obtain samples of $k,\xi$ from the posterior distribution \eqref{eq:posterior} or \eqref{eq:approx}. Due to the tensor product structure, the sampling procedure can be performed in the component individually. Suppose $x_i$ is the current update component. When modifying $x_i$, other components are invariant. We design three transition strategies to traverse the state space: (1) Birth: add a knot with the probability $b_{k_i}=c\min (1,\{(n_i-k_i)/(k_i+1)\}^{1-\gamma})$; (2) Death: delete a knot with the probability $d_{k_i}=c\min (1, \{k_i/(n_i-k_i+1)\}^{1-\gamma})$; (3) Relocation: change the position of one knot with the probability $r_{k_i}=1-b_{k_i}-d_{k_i}$. The hyper-parameter $c$ is of the interval $(0,0.5)$. Notably, $\pi(k_i)b_{k_i}=\pi(k_i+1)d_{k_i+1}$, satisfying the detailed balance equation for the prior of the knot number. Suppose the jumping probability is $q(k_i',\xi_i'|k_i,\xi_i)$. Let $\xi_{i,k_i}\subset \eta_i$ be the locations of $k_i$ knots in the $i$-th component. The concrete proposal distributions are specified in the following manner:
\begin{enumerate}
    \item \emph{Birth step}. Select a knot from the remaining candidate knots $\eta_i \backslash \xi_{i,k_i}$ uniformly, and add it into $\xi_{i,k_i}$. Then $q(k_i+1,\xi_{i,k_i+1}|k_i,\xi_{i,k_i})=b_{k_i}/(n_i-k_i)$.
    \item \emph{Death step}. Select a knot from the current knots $\xi_{i,k_i}$ uniformly, and delete it. Then $q(k_i-1,\xi_{i,k_i-1}|k_i,\xi_{i,k_i})=d_{k_i}/k_i$.
    \item \emph{Relocation step}. Select a knot from $\xi_{i,k_i}$ and another knot from $\eta_i\backslash \xi_{i,k_i}$ uniformly. Exchange their positions. Then $q(k_i,\xi_{i,k_i}' | k_i,\xi_{i,k_i})=r_{k_i}/\{k_i(n_i-k_i)\}$.
\end{enumerate}

\iffalse
\emph{Birth step}. Select a knot from the remaining candidate knots $\eta_i \backslash \xi_{i,k_i}$ uniformly, and add it into $\xi_{i,k_i}$. Then $q(k_i+1,\xi_{i,k_i+1}|k_i,\xi_{i,k_i})=b_{k_i}/(n_i-k_i)$.

\emph{Death step}. Select a knot from the current knots $\xi_{i,k_i}$ uniformly, and delete it. Then $q(k_i-1,\xi_{i,k_i-1}|k_i,\xi_{i,k_i})=d_{k_i}/k_i$.

\emph{Relocation step}. Select a knot from $\xi_{i,k_i}$ and another knot from $\eta_i\backslash \xi_{i,k_i}$ uniformly. Exchange their positions. Then $q(k_i,\xi_{i,k_i}' | k_i,\xi_{i,k_i})=r_{k_i}/\{k_i(n_i-k_i)\}$.
\fi

Assume that $(k,\xi)$ is the current status and $(k',\xi')$ is the candidate status. To sample from the target posterior distribution, the Metropolis-Hastings algorithm implies that,
\begin{equation}
    \label{eq:detailed_balance}
    p(k,\xi|y)q(k',\xi'|k,\xi)\alpha(k',\xi'|k,\xi) = p(k',\xi'|y)q(k,\xi|k',\xi')\alpha(k,\xi|k',\xi'),
\end{equation}
where $\alpha(k',\xi'|k,\xi),\alpha(k,\xi|k',\xi')$ are the acceptance probabilities. %Besides, the proposal distribution $q(k',\xi'|k,\xi)$ is $q(k_i',\xi_i'|k_i,\xi_i)$ when updating knots in the $i$-th component.

\begin{lemma}
\label{lm:accept}
    With the specified proposal distribution and the posterior density of \eqref{eq:detailed_balance}, the acceptance probability and its EBIC approximation in RJMCMC are
    \begin{equation}
        \label{eq:accept}
        \begin{split}
            \alpha(k',\xi'|k,\xi) & = \min\{1, ~(m+1)^{(\nu-\nu')/2}(a_{k,\xi}/a_{k',\xi'})^{m/2}\},\\
            \hat{\alpha}(k',\xi'|k,\xi) & = \min \{1, ~ m^{(\nu-\nu')/2} ( \hat{\sigma}^2/(\hat{\sigma}')^2 )^{m/2} \},\\
        \end{split}
    \end{equation}
    where $\nu,\nu'$ are the dimensions of the spline space.
\end{lemma}

Lemma \ref{lm:accept} implies that $\sqrt{m+1},\sqrt{m}$ are the dimensional penalty factors for the likelihood. %This coincides to the method of \citet{10.1093/biomet/88.4.1055}. 
Comparing $\alpha(k',\xi'|k,\xi)$ and $\hat{\alpha}(k',\xi'|k,\xi)$ in \eqref{eq:accept}, $\alpha\approx \hat{\alpha}$ when the sample size is sufficiently large. The overall extended Bayesian adaptive regression spline approach is listed as Algorithm \ref{alg:ebars}.

\begin{algorithm}
\caption{Extended Bayesian adaptive regression spline algorithm via RJMCMC}
\label{alg:ebars}
\begin{algorithmic}[1]
\Statex \emph{Input:} Labeled observations $\{(x_i,y_i)\}_{i=1}^m$, hyper-parameters $0\leq \gamma\leq 1$ and $0<c<0.5$, jumping steps $I$, the number of candidate knots $n$.
\State Create the candidate knots $\eta$. Initialize the starting knots $(k^{(0)},\xi^{(0)})$.
\For{$i=0,\ldots,I-1$}
\State Determine the $j$-th component to update randomly and choose the transition strategy.
% \State Calculate $b_{k^{(i)}_j},d_{k^{(i)}_j},r_{k^{(i)}_j}$ to choose the transition strategy.
\State Solve $(k',\xi')$ for the next movement by the proposal distribution.
\State Compute $\alpha(k',\xi'|k^{(i)},\xi^{(i)})$ by \eqref{eq:accept}. Sample $u\sim U(0,1)$.
\If{$u<\alpha(k',\xi'|k^{(i)},\xi^{(i)})$} 
\State Update the knots by $(k^{(i+1)},\xi^{(i+1)})=(k',\xi')$;
\Else
\State Return to Step 3 and repeat.
\EndIf
\EndFor
%\State Calculate the maximum likelihood estimators $\hat{\beta}^{(I)},\hat{\sigma}^{(I)}$ based on \eqref{eq:spline_regression}.
%\Statex \emph{Output:} The estimators $k^{(I)},\xi^{(I)},\hat{\beta}^{(I)},\hat{\sigma}^{(I)}$.
\Statex \emph{Output:} The posterior samples $\{k^{(i)},\xi^{(i)}\}_{i=1}^I$.
\end{algorithmic}
\end{algorithm}
\section{Experiment}

We conduct EBARS in the knot inference and manifold denoising. For the knot inference, the performance is compared with the segmented method of \citet{muggeo2003estimating} and the modified maximum likelihood (MML) method of \citet{doi:10.1080/01621459.2021.1947307}. %We calculate the absolute errors of the estimated knots for evaluation. 
For the manifold denoising, the performance is compared with manifold fitting under unbounded noise (MFUN) of \citet{yao2023manifoldfitting}, putative manifold fitting (PMF) of \citet{pmlr-v75-fefferman18a}, principal manifold estimation (PME) of \citet{10.1111/rssb.12416} and principal curves (PC) of \citet{principalcurve}. %We calculate the geometric mean squared distance for evaluation. 
Numerical experiments show that the proposed method performs well in finite samples of all scenarios.

\subsection{Knot inference}

\begin{figure}
    \centering
    \includegraphics[width=0.9\textwidth]{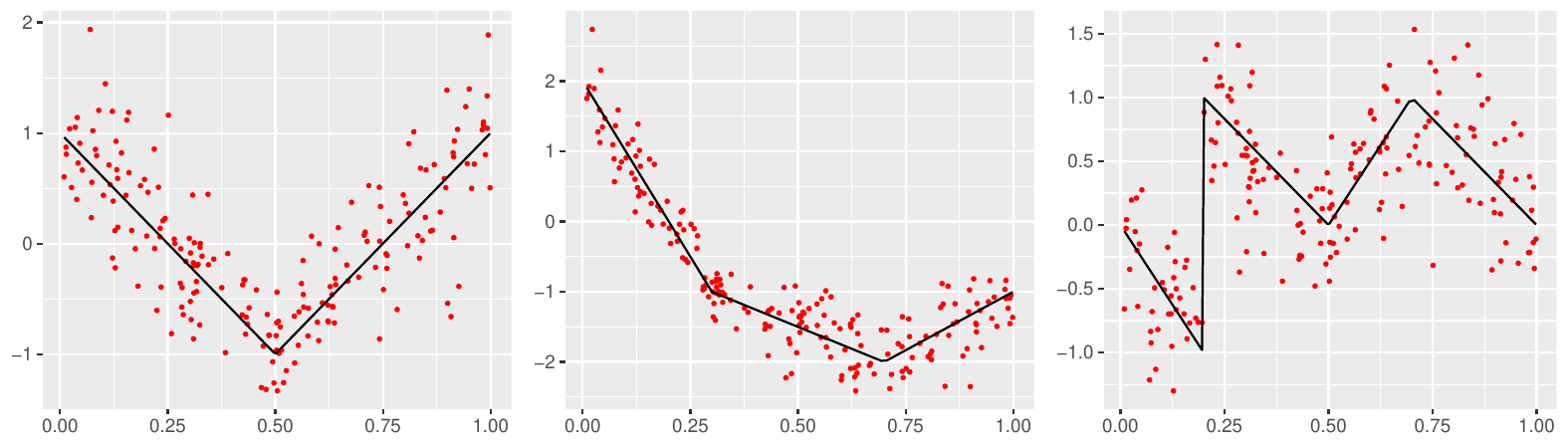}
    \caption{Linear splines with one, two and four knots. Black segments are true spline curves and red points are noisy observed data with $m=200$.}
    \label{fig:knot_estimation}
\end{figure}

%In many problems, the quantities of main interest are change-points or break-points in the model, such as change-points detection in time series \citep{aminikhanghahi2017survey,truong2020selective}. Thus researchers want to construct an accurate estimation for change-points with rigorous statistical approaches \citep{doi:10.1080/01621459.2015.1073154,doi:10.1080/01621459.2021.1947307}. 
We perform EBARS for knot inference in linear spline regression. % The linear spline is a piecewise linear model. 
Line segments are connected in unique knots and the spline is discontinuous at the location where two knots coincide. The data is generated from three linear spline models with one, two and four knots as shown in Figure \ref{fig:knot_estimation}, where the knot locations are respectively $(0.5),~(0.3,0.7),~(0.2,0.2,0.5,0.7)$. The functions are continuous except for the spline with $k=4$. The noise is Gaussian with standard deviation $0.4$, $0.3$, $0.4$. 

\begin{figure}
    \centering
    \includegraphics[width=0.9\textwidth]{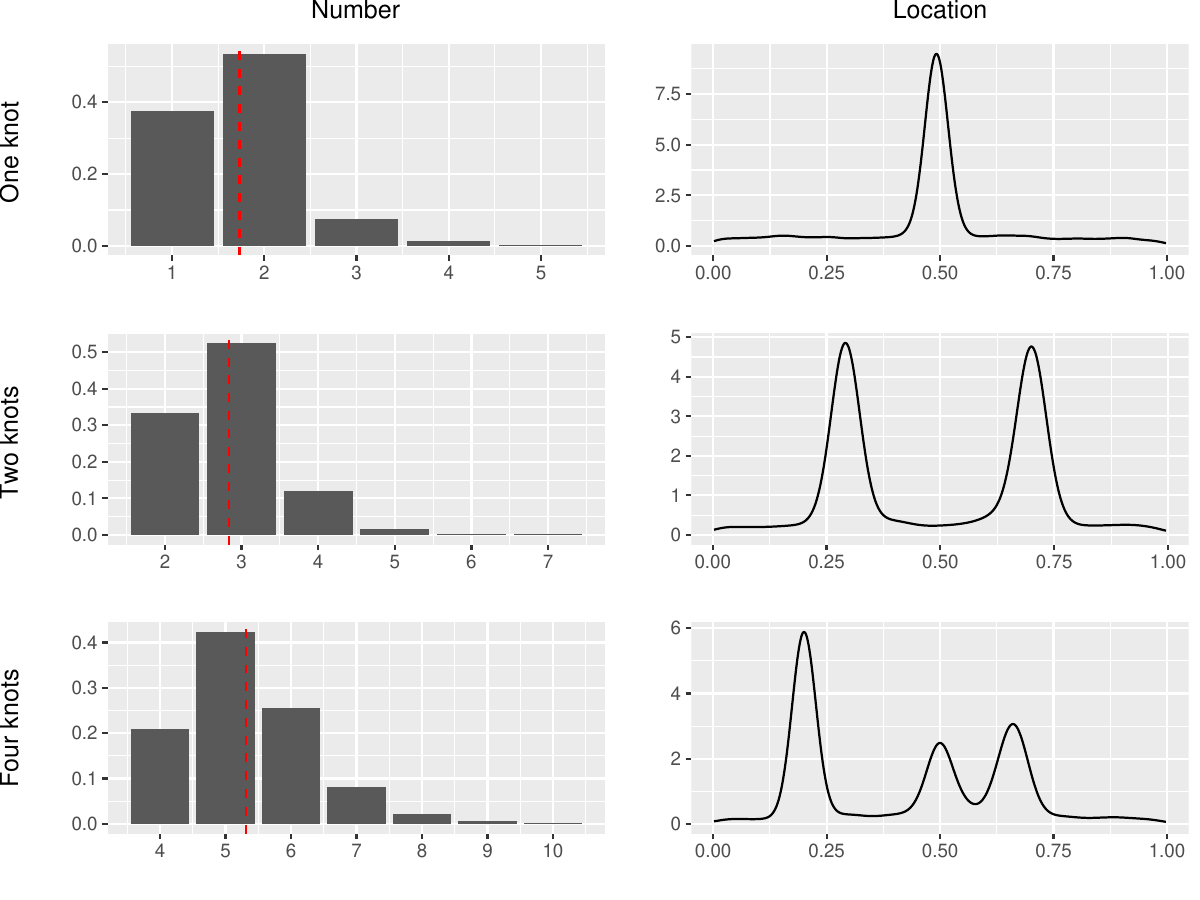}
    \caption{The posterior distributions of knots in three scenarios by EBARS. Left channels are the histograms of the knot number, where red dashed vertical lines indicate the mean number. Right channels are the posterior density plots of the knot location.}
    \label{fig:knot_inference}
\end{figure}

Figure \ref{fig:knot_inference} illustrates the posterior distribution of the knot number and location by EBARS with $\gamma=1$ in $k=1,2,4$. The sample size is $500$ in all scenarios. We simulate $5000$ samples of $(k,\xi)$ via RJMCMC after $5000$ burning steps. Histograms show that the average knot number is close to the true value with negligible error. From the density plots, the posterior mass concentrates on the correct location of change points in all cases even though the knot number is unknown. Especially, for $k=4$, the posterior density at $0.2$ is double at $0.5$ and $0.7$, meaning that $0.2$ is chosen as the knot twice.

To evaluate the performance of knot inference, we compare EBARS with Segmented of Muggeo and MML of Guangyu Yang and Zhang. In this experiment, the knot number is given since Segmented and MML cannot estimate $k$. The absolute error of knot location is calculated as criteria. Three methods are evaluated under sample sizes $m=200,500$ and the experiment is repeated $50$ times.

Simulation results of the mean and standard deviation of absolute errors are summarized in Tables \ref{tab:knot_1} and \ref{tab:knot_2}. When $k=1,2$, three methods obtain fairly small absolute errors. However, when $k=4$, the absolute errors of EBARS are much less than other two algorithms for all knots. MML struggles to estimate the duplicate knots $0.2$ and causes horrible bias. Segmented outperforms MML but it is still much less accurate than EBARS. Compared with MML and Segmented, our EBARS is robust and attain a precise estimation with jumping discontinuity and increasing number of knots. Furthermore, MML and Segmented are exclusively applied in linear splines and they require the knot number to be given, whereas EBARS is convenient to estimate the knot number and location simultaneously in splines of any degree.

\begin{table}
\caption{Absolute errors for knot location estimation of three methods in $k=1,2$}
\label{tab:knot_1}
\centering
\begin{tabular}{ccccc}
\toprule
\multirow{2}{*}{Methods} & \multirow{2}{*}{m}   & One knot       & \multicolumn{2}{c}{Two knots}    \\
\cline{3-5}
                         &                      & Knot $1$         & Knot 1         & Knot 2         \\
                         \midrule
EBARS                    & \multirow{3}{*}{200} & 0.0219(0.0126) & 0.0167(0.0150)  & 0.0248(0.0257) \\
MML                      &                      & 0.0163(0.0099) & 0.0138(0.0118) & 0.0170(0.0214)  \\
Segmented                &                      & 0.0159(0.0102) & 0.0117(0.0113) & 0.0186(0.0204) \\
\midrule
EBARS                    & \multirow{3}{*}{500} & 0.0100(0.0076)  & 0.0095(0.0065) & 0.0142(0.0122) \\
MML                      &                      & 0.0075(0.0060)  & 0.0068(0.0058) & 0.0102(0.0076) \\
Segmented                &                      & 0.0079(0.0061) & 0.0070(0.0059)  & 0.0113(0.0081) \\
\bottomrule

\end{tabular}
\end{table}

\begin{table}
\caption{Absolute errors for knot location estimation of three methods in $k=4$}
\label{tab:knot_2}
\centering
\begin{tabular}{cccccc}
\toprule
\multirow{2}{*}{Methods} & \multirow{2}{*}{m}   & \multicolumn{4}{c}{Four knots}                                      \\
\cline{3-6}
                         &                      & Knot 1         & Knot 2         & Knot 3         & Knot 4         \\
                         \midrule
EBARS                    & \multirow{3}{*}{200} & 0.0058(0.0060) & 0.0044(0.0043) & 0.0206(0.0199) & 0.0170(0.0096) \\
MML                      &                      & 0.3437(1.1010) & 0.1163(0.1283) & 0.0978(0.1006) & 0.1012(0.1454) \\
Segmented                &                      & 0.0366(0.0494) & 0.0712(0.1067) & 0.0799(0.0889) & 0.0752(0.0805) \\
\midrule
EBARS                    & \multirow{3}{*}{500} & 0.0025(0.0029) & 0.0026(0.0025) & 0.0194(0.0172) & 0.0150(0.0137) \\
MML                      &                      & 0.2737(0.4266) & 0.1725(0.1683) & 0.1065(0.1800) & 0.2131(0.5211) \\
Segmented                &                      & 0.0171(0.0346) & 0.0283(0.0775) & 0.0395(0.0825) & 0.0349(0.0805) \\
\bottomrule
\end{tabular}
\end{table}
% \input{Experiments/surface_regression}

%\section{Application}
\subsection{Manifold denoising}

With the assumption of linearity, principal component analysis is a prevailing and efficient method for dimension reduction in high-dimensional space. However, the simple method poses a limitation to respect the nonlinear relationship. 
% Thus, it is essential to develop a powerful and sophisticated approach for nonlinear data analysis. 
Manifold estimation is a technique for modelling the complicated dependence in high-dimensional data, denoising the observations and estimating the low-dimensional latent manifold. Refer to \citet{doi:10.1073/pnas.2311436121} for a detailed review about manifold estimation. 
%Many statistical models have been established to achieve the objective, such as CycleGAN of \citet{doi:10.1073/pnas.2311436121} and principal manifold estimation (PME) of \citet{10.1111/rssb.12416}.

Assume that the observed data $\{x_i\}_{i=1}^m$ is generated from a low-dimensional latent manifold of a high-dimensional ambient space with random noise. That is,
\begin{equation*}
    X=W+\epsilon, \quad X,\epsilon\in \mathbb{R}^D,~W\in \mathbb{M}^d,
\end{equation*}
where $\mathbb{M}^d$ is a $d$-dimensional submanifold of $\mathbb{R}^D$ with $d\leq D$, $W$ is generated from a probability distribution supported in $\mathbb{M}^d$, and $\epsilon$ is $D$-dimensional noise independent of $W$. %The main objective of the manifold estimation is to fit the latent manifold by using $\{x_i\}_{i=1}^m$. According to differentiable geometry, a $d$-dimensional Riemannian manifold with global parametrization is determined by an exponential map $f=(f_1,\ldots,f_D):\mathbb{R}^d\rightarrow \mathbb{M}^d\subset \mathbb{R}^D$, where $\mathbb{R}^d$ is the corresponding tangent space. Therefore, the manifold estimation is also equivalent to finding such a parametrization.
In this article, we propose a two stage manifold estimation (TSME) technique for modelling $\mathbb{M}^d$. The primary challenge of parametrization is attributed to the lack of the required pairs of training data, \emph{i.e.}, (predictor, response). %Thus, it is an unsupervised learning problem. 
We resolve the issue simply by combining the manifold embedding and reconstruction. The manifold embedding is a nonlinear dimensional reduction method. Many impressive algorithms have been proposed in the literature, such as ISOMAP \citep{balasubramanian2002isomap}, Laplacian eigenmaps \citep{belkin2003laplacian} and UMAP \citep{mcinnes2018umap}. 

At the first stage, we apply the manifold embedding technique in $\{x_i\}_{i=1}^m$ given the intrinsic dimension $d$, yielding a projection map $\hat{g}:\mathbb{R}^D\rightarrow \mathbb{R}^d$. Let $u_i=\hat{g}(x_i)$ for $i=1,\ldots,m$. Then $\{(u_i,x_i)\}_{i=1}^m$ are the training samples for the manifold reconstruction. At the second stage, we execute the regression in $\{(u_i,x_i)\}_{i=1}^m$ to reconstruct the manifold. For example, we can use the proposed EBARS method to estimate the reconstruction map $\hat{f}$. Due to the automatic knot selection, EBARS is eligible to model the complex relationships between $u$ and $x$. Consequently, the manifold estimation is completed by the composition of the embedding $\hat{g}$ and the reconstruction $\hat{f}$. Especially, $\{\hat{f}\circ \hat{g}(x_i)\}_{i=1}^m$ can serve as the denoised representation of $\{x_i\}_{i=1}^m$. %Furthermore, the composition $\hat{f}\circ\hat{g}$ can be considered as a solution to the minimum problem of the loss function $\mathbb{E}\left|\left| X-f\circ g(X) \right|\right|^2$.

\begin{figure}
    \centering
    \includegraphics[width=0.9\textwidth]{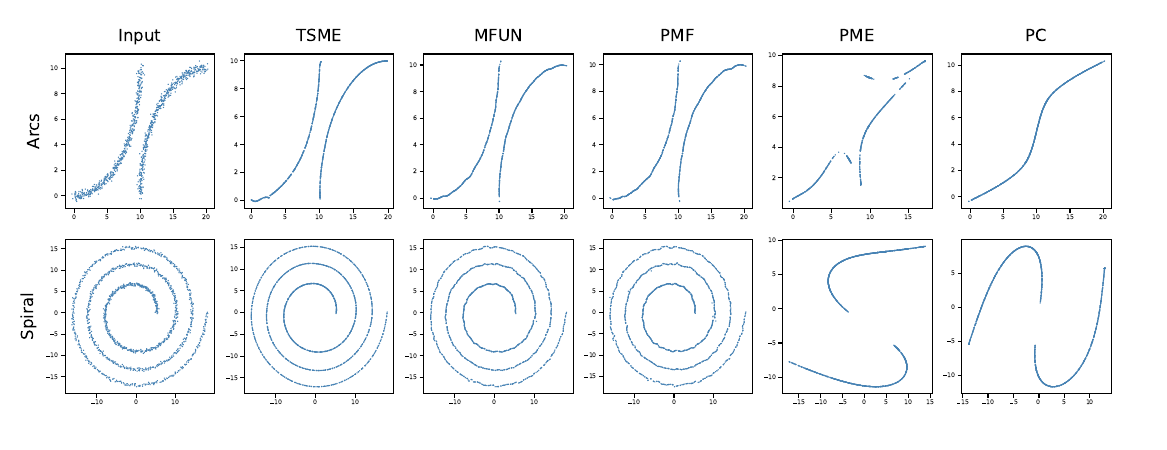}
    \caption{Manifold denoise results for two curve scenarios. Each plot indicates the training samples or the denoised representation obtained by our TSME, MFUN, PMF, PME and PC.}
    \label{fig:d1}
\end{figure}

\begin{figure}
    \centering
    \includegraphics[width=0.95\textwidth]{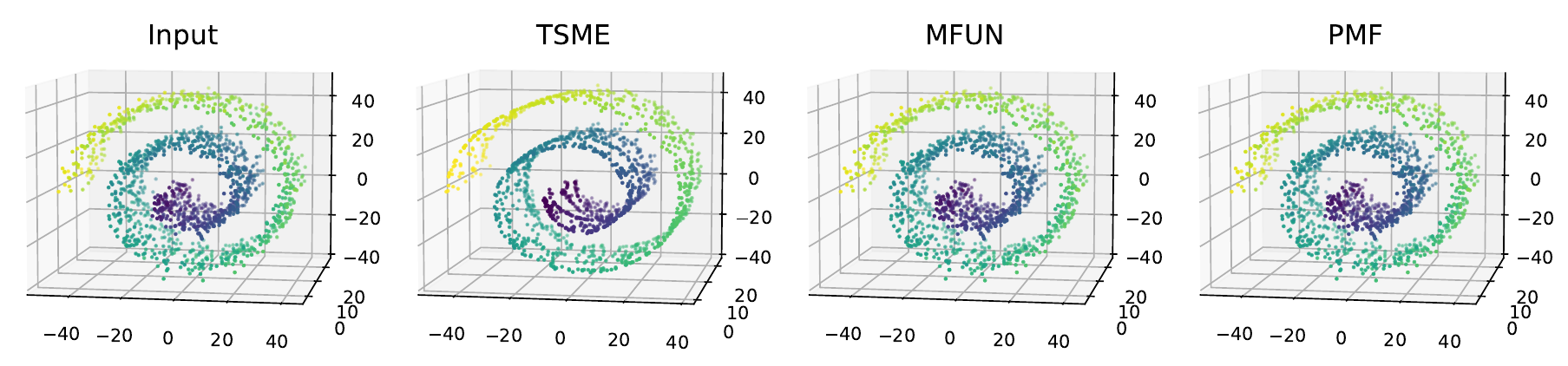}
    \caption{Manifold denoise results for the Swiss roll. Each plot indicates the training samples or the denoised representation obtained by our TSME, MFUN and PMF.}
    \label{fig:d2}
\end{figure}

We conduct the TSME method in manifold denoising and compare its performance with MFUN, PMF, PME and PC. Notably, PME and PC adopt a similar framework with our TSME except that they utilize smoothing splines for reconstruction. The data is generated from two disconnected arcs in $\mathbb{R}^2$, a spiral curve in $\mathbb{R}^2$ and a Swiss roll in $\mathbb{R}^3$, as shown in Figures \ref{fig:d1} and \ref{fig:d2}.%corresponding to $(d=1,D=2)$ and $(d=2,D=3)$. 
ISOMAP is used as the embedding map of TSME and the initial projection of PME in the spiral and  Swiss roll. For the arc case, we substitute UMAP for ISOMAP due to the disconnectivity. Besides, for the Swiss roll, we only compare TSME, MFUN and PMF, as PC is limited to the curve case and PME is very slow with disastrous performance. The random noise is Gaussian with standard deviation $0.2,0.2,1.5$.

Simulation results including the training samples and the denoised data are visualized in Figures \ref{fig:d1} and \ref{fig:d2}. The sample sizes are $m=1000,1500,3000$. Compared with TSME, the PME and PC methods collapse dramatically in curve cases. In arcs, the disconnectivity of $\mathbb{M}^d$ causes the jumping discontinuity of the reconstruction map. %As mentioned above, EBARS is superior to the smoothing splines in describing sharp changes. 
In spiral, the dependency is complex due to the intricate manifold, posing a critical challenge in selecting the optimal penalty weight of smoothing splines. %And the smoothness in EBARS is controlled based on the number of knots without the penalty term. 
For MFUN and PMF, the denoised representations remain corrupted in the spiral and Swiss roll cases. They don't mitigate the random noise successfully and fail to estimate the latent manifolds accurately.

To evaluate the performance quantitatively, we define the geometric mean squared distance as
\begin{equation*}
    \text{GMSD}=\frac{1}{m}\sum_{i=1}^m \text{dist}(\hat{f}\circ\hat{g}(x_i),\mathbb{M}^d)^2,
\end{equation*}
where $\text{dist}(x,\mathbb{M}^d)=\inf_{x'\in\mathbb{M}^d}||x-x'||$. The GMSD measures the corrupted extent of the data away from the manifold. Table \ref{tab:gmsd} illustrates the mean and standard deviation of GMSD in three manifolds of $20$ duplicates. TSME obtains significantly smaller GMSD than the training samples in all scenarios and achieves the smallest GMSD in most cases, demonstrating that our method contributes to excellent manifold denoising. For MFUN and PMF, the performance in the $d=1$ cases is satisfactory but the noise is not reduced in the Swiss roll. PME and PC are completely incapable of estimating the latent manifolds. %This indicates again the great advantage of EBARS over the smoothing spline with equally-spaced fixed knots.

\begin{table}
    \centering
    \caption{Summary of GMSD with respect to five methods in three manifolds}
    \label{tab:gmsd}
    \begin{tabular}{@{}cccc@{}}
    \toprule
          & Arcs                    & Spiral                  & Swiss roll              \\ \midrule
    $m$ & 500 & 1000 & 1500 \\
    Input & 0.0404(0.0020)          & 0.0397(0.0021)          & 2.6033(0.0729)          \\
    TSME  & 0.0036(0.0021)          & 0.0019(0.0003) & 0.5299(0.0653) \\
    MFUN  & 0.0039(0.0011) & 0.0112(0.0011)          & 2.5549(0.0735)          \\
    PMF   & 0.0051(0.0011)          & 0.0165(0.0017)          & 2.5994(0.0733)          \\
    PME   & 0.4790(0.2585)          & 2618.6(9887.4)          &            $\times$             \\
    PC    & 0.7661(0.0077)          & 2.9539(0.0633)          &            $\times$             \\ \midrule
        $m$ & 1000 & 1500 & 3000 \\
    Input & 0.0399(0.0013)          & 0.0401(0.0013)          & 2.6423(0.0621)          \\
    TSME  & 0.0044(0.0033)          & 0.0015(0.0002) & 0.5004(0.0605) \\
    MFUN  & 0.0018(0.0004) & 0.0078(0.0005)          & 2.5377(0.0629)          \\
    PMF   & 0.0023(0.0007)          & 0.0113(0.0008)          & 2.6241(0.0625)          \\
    PME   & 0.3135(0.1353)          & 151.87(449.52)          &            $\times$             \\
    PC    & 0.7589(0.0072)          & 2.9601(0.0699)          &            $\times$             \\ 
    \bottomrule
\end{tabular}
\end{table}

\section{Conclusion}

This article is motivated by the knot selection problem in the multivariate spline regression. %We take into account the complexity of the model space and specify a prior on the number of knots to hedge the complexity. 
The proposed method estimates the number and location of spline knots simultaneously and accurately. %regardless of the true number of knots and the sample size. 
It provides a mechanism for easy interpretation and function fitting with jumping discontinuity. The trade-off for advantages is the additional computational cost, attributed to RJMCMC. Nevertheless, it will not cause a serious problem in small or moderately large datasets.

There are several potential jobs in the future. We intend to establish a theoretical framework of the algorithm including the consistency and convergence rate. The primary challenge is how to define an appropriate distance between parameters with varying dimensions. Furthermore, the domain space is required to be a cube. We aim for an extension in the general domain. A possible solution is substituting a flexible multivariate spline for the tensor product spline.
\section{Code and technical appendices}

The source package EBARS is accessible on the GitHub repository \url{https://github.com/junhuihe2000/EBARS} and R codes of examples in Section 4 are available on the GitHub repository \url{https://github.com/junhuihe2000/exampleEBARS}. The technical appendices include proofs of Lemmas 1 and 3, and additional simulations in curve and surface fitting.
\subsubsection*{Acknowledgements}

% We sincerely thank the associate editor and the referees for their comments to improve the quality of the paper.

Yang's research was partially supported by the National Natural Science Foundation of China grant (12271286, 11931001). Kang's research was partially supported by the following grants: NIH R01DA048993, NIH R01MH105561 and NSF IIS2123777.

%USE THE BELOW OPTIONS IN CASE YOU NEED AUTHOR YEAR FORMAT.
{
%\small
\bibliographystyle{abbrvnat}
\bibliography{refs/ref}

\begin{thebibliography}{36}
\providecommand{\natexlab}[1]{#1}
\providecommand{\url}[1]{\texttt{#1}}
\expandafter\ifx\csname urlstyle\endcsname\relax
  \providecommand{\doi}[1]{doi: #1}\else
  \providecommand{\doi}{doi: \begingroup \urlstyle{rm}\Url}\fi

\bibitem[Aminikhanghahi and Cook(2017)]{aminikhanghahi2017survey}
S.~Aminikhanghahi and D.~J. Cook.
\newblock A survey of methods for time series change point detection.
\newblock \emph{Knowledge and Information Systems}, 51\penalty0 (2):\penalty0 339--367, 2017.

\bibitem[Balasubramanian and Schwartz(2002)]{balasubramanian2002isomap}
M.~Balasubramanian and E.~L. Schwartz.
\newblock The isomap algorithm and topological stability.
\newblock \emph{Science}, 295\penalty0 (5552):\penalty0 7--7, 2002.

\bibitem[Belkin and Niyogi(2003)]{belkin2003laplacian}
M.~Belkin and P.~Niyogi.
\newblock Laplacian eigenmaps for dimensionality reduction and data representation.
\newblock \emph{Neural Computation}, 15\penalty0 (6):\penalty0 1373--1396, 2003.

\bibitem[Bolton and Heard(2018)]{doi:10.1080/01621459.2018.1423984}
A.~D. Bolton and N.~A. Heard.
\newblock Malware family discovery using reversible jump {MCMC} sampling of regimes.
\newblock \emph{Journal of the American Statistical Association}, 113\penalty0 (524):\penalty0 1490--1502, 2018.

\bibitem[Chapple et~al.(2020)Chapple, Peak, and Hemal]{https://doi.org/10.1002/sim.8511}
A.~G. Chapple, T.~Peak, and A.~Hemal.
\newblock A novel {B}ayesian continuous piecewise linear log-hazard model, with estimation and inference via reversible jump {M}arkov chain {M}onte {C}arlo.
\newblock \emph{Statistics in Medicine}, 39\penalty0 (12):\penalty0 1766--1780, 2020.

\bibitem[Chen et~al.(2011)Chen, Chan, Gerlach, and Hsieh]{chen2011comparison}
C.~W. Chen, J.~S. Chan, R.~Gerlach, and W.~Y. Hsieh.
\newblock A comparison of estimators for regression models with change points.
\newblock \emph{Statistics and Computing}, 21:\penalty0 395--414, 2011.

\bibitem[Chen and Chen(2008)]{10.1093/biomet/asn034}
J.~Chen and Z.~Chen.
\newblock {Extended Bayesian information criteria for model selection with large model spaces}.
\newblock \emph{Biometrika}, 95\penalty0 (3):\penalty0 759--771, 09 2008.

\bibitem[Chen and Chen(2012)]{bb6df906-631b-39ae-98e5-1c3a3cd1049c}
J.~Chen and Z.~Chen.
\newblock Extended {BIC} for small-n-large-p sparse {GLM}.
\newblock \emph{Statistica Sinica}, 22\penalty0 (2):\penalty0 555--574, 2012.

\bibitem[Denison et~al.(1998{\natexlab{a}})Denison, Mallick, and Smith]{Denison1998MARS}
D.~G.~T. Denison, B.~K. Mallick, and A.~F.~M. Smith.
\newblock Bayesian {MARS}.
\newblock \emph{Statistics and Computing}, 8:\penalty0 337--346, 1998{\natexlab{a}}.

\bibitem[Denison et~al.(1998{\natexlab{b}})Denison, Mallick, and Smith]{https://doi.org/10.1111/1467-9868.00128}
D.~G.~T. Denison, B.~K. Mallick, and A.~F.~M. Smith.
\newblock Automatic {B}ayesian curve fitting.
\newblock \emph{Journal of the Royal Statistical Society: Series B (Statistical Methodology)}, 60\penalty0 (2):\penalty0 333--350, 1998{\natexlab{b}}.

\bibitem[Dierckx(1995)]{dierckx1995curve}
P.~Dierckx.
\newblock \emph{Curve and Surface Fitting with Splines}.
\newblock Oxford University Press, New York, 1995.

\bibitem[Dimatteo et~al.(2001)Dimatteo, Genovese, and Kass]{10.1093/biomet/88.4.1055}
I.~Dimatteo, C.~R. Genovese, and R.~E. Kass.
\newblock {Bayesian curve‐fitting with free‐knot splines}.
\newblock \emph{Biometrika}, 88\penalty0 (4):\penalty0 1055--1071, 12 2001.
\newblock ISSN 0006-3444.

\bibitem[Fearnhead(2006)]{fearnhead2006exact}
P.~Fearnhead.
\newblock Exact and efficient {B}ayesian inference for multiple changepoint problems.
\newblock \emph{Statistics and Computing}, 16:\penalty0 203--213, 2006.

\bibitem[Fefferman et~al.(2018)Fefferman, Ivanov, Kurylev, Lassas, and Narayanan]{pmlr-v75-fefferman18a}
C.~Fefferman, S.~Ivanov, Y.~Kurylev, M.~Lassas, and H.~Narayanan.
\newblock Fitting a putative manifold to noisy data.
\newblock In S.~Bubeck, V.~Perchet, and P.~Rigollet, editors, \emph{Proceedings of the 31st Conference On Learning Theory}, volume~75, pages 688--720. PMLR, 06--09 Jul 2018.

\bibitem[Foygel and Drton(2010)]{NIPS2010_072b030b}
R.~Foygel and M.~Drton.
\newblock Extended {B}ayesian information criteria for {G}aussian graphical models.
\newblock \emph{Advances in Neural Information Processing Systems}, 23, 2010.

\bibitem[Green(1995)]{10.1093/biomet/82.4.711}
P.~J. Green.
\newblock {Reversible jump Markov chain Monte Carlo computation and Bayesian model determination}.
\newblock \emph{Biometrika}, 82\penalty0 (4):\penalty0 711--732, 12 1995.

\bibitem[Green and Silverman(1994)]{green1994nonparametric}
P.~J. Green and B.~W. Silverman.
\newblock \emph{Nonparametric Regression and Generalized Linear Models: A Roughness Penalty Approach}.
\newblock Chapman and Hall, London, 1994.

\bibitem[Gu(2013)]{gu2013smoothing}
C.~Gu.
\newblock \emph{Smoothing Spline {ANOVA} Models}.
\newblock Springer, New York, 2013.

\bibitem[Guangyu~Yang and Zhang(2023)]{doi:10.1080/01621459.2021.1947307}
B.~Z. Guangyu~Yang and M.~Zhang.
\newblock Estimation of knots in linear spline models.
\newblock \emph{Journal of the American Statistical Association}, 118\penalty0 (541):\penalty0 639--650, 2023.

\bibitem[Hastie and Stuetzle(1989)]{principalcurve}
T.~Hastie and W.~Stuetzle.
\newblock Principal curves.
\newblock \emph{Journal of the American Statistical Association}, 84\penalty0 (406):\penalty0 502--516, 1989.

\bibitem[Kass and Wasserman(1995)]{e447aec4-501c-3132-a415-1b260054b4a8}
R.~E. Kass and L.~Wasserman.
\newblock A reference {B}ayesian test for nested hypotheses and its relationship to the {S}chwarz criterion.
\newblock \emph{Journal of the American Statistical Association}, 90\penalty0 (431):\penalty0 928--934, 1995.

\bibitem[Lerman(1980)]{10.2307/2346413}
P.~M. Lerman.
\newblock Fitting segmented regression models by grid search.
\newblock \emph{Journal of the Royal Statistical Society Series C: Applied Statistics}, 29\penalty0 (1):\penalty0 77--84, 3 1980.

\bibitem[Luo et~al.(2015)Luo, Xu, and Chen]{Luo2015}
S.~Luo, J.~Xu, and Z.~Chen.
\newblock {Extended Bayesian information criterion in the Cox model with a high-dimensional feature space}.
\newblock \emph{Annals of the Institute of Statistical Mathematics}, 67:\penalty0 287–311, 2015.

\bibitem[Marsh and Cormier(2001)]{marsh2001spline}
L.~C. Marsh and D.~R. Cormier.
\newblock \emph{Spline regression models}.
\newblock Sage, Iowa, 2001.

\bibitem[McInnes et~al.(2020)McInnes, Healy, and Melville]{mcinnes2018umap}
L.~McInnes, J.~Healy, and J.~Melville.
\newblock {UMAP}: Uniform manifold approximation and projection for dimension reduction.
\newblock \emph{Preprint arXiv:1802.03426}, 2020.

\bibitem[Meng and Eloyan(2021)]{10.1111/rssb.12416}
K.~Meng and A.~Eloyan.
\newblock Principal manifold estimation via model complexity selection.
\newblock \emph{Journal of the Royal Statistical Society Series B: Statistical Methodology}, 83\penalty0 (2):\penalty0 369--394, 03 2021.

\bibitem[Muggeo(2003)]{muggeo2003estimating}
V.~M. Muggeo.
\newblock Estimating regression models with unknown break-points.
\newblock \emph{Statistics in Medicine}, 22\penalty0 (19):\penalty0 3055--3071, 2003.

\bibitem[Muggeo(2008)]{muggeo2008segmented}
V.~M. Muggeo.
\newblock Segmented: an {R} package to fit regression models with broken-line relationships.
\newblock \emph{R news}, 8\penalty0 (1):\penalty0 20--25, 2008.

\bibitem[Perperoglou et~al.(2019)Perperoglou, Sauerbrei, Abrahamowicz, and Schmid]{perperoglou2019review}
A.~Perperoglou, W.~Sauerbrei, M.~Abrahamowicz, and M.~Schmid.
\newblock A review of spline function procedures in {R}.
\newblock \emph{BMC Medical Research Methodology}, 19:\penalty0 1--16, 2019.

\bibitem[Ritabrata~Das and Zheng(2016)]{doi:10.1080/01621459.2015.1073154}
B.~N. Ritabrata~Das, Moulinath~Banerjee and H.~Zheng.
\newblock Fast estimation of regression parameters in a broken-stick model for longitudinal data.
\newblock \emph{Journal of the American Statistical Association}, 111\penalty0 (515):\penalty0 1132--1143, 2016.

\bibitem[Schumaker(2007)]{schumaker2007spline}
L.~Schumaker.
\newblock \emph{Spline Functions: Basic Theory}.
\newblock Cambridge University Press, New York, 2007.

\bibitem[Truong et~al.(2020)Truong, Oudre, and Vayatis]{truong2020selective}
C.~Truong, L.~Oudre, and N.~Vayatis.
\newblock Selective review of offline change point detection methods.
\newblock \emph{Signal Processing}, 167:\penalty0 107299, 2020.

\bibitem[Wahba(1990)]{wahba1990spline}
G.~Wahba.
\newblock \emph{Spline Models for Observational Data}.
\newblock Society for Industrial and Applied Mathematics, Philadelphia, 1990.

\bibitem[Wood(2003)]{10.1111/1467-9868.00374}
S.~N. Wood.
\newblock {Thin Plate Regression Splines}.
\newblock \emph{Journal of the Royal Statistical Society Series B: Statistical Methodology}, 65\penalty0 (1):\penalty0 95--114, 2003.

\bibitem[Yao and Xia(2023)]{yao2023manifoldfitting}
Z.~Yao and Y.~Xia.
\newblock Manifold fitting under unbounded noise.
\newblock \emph{Preprint arXiv:1909.10228}, 2023.

\bibitem[Yao et~al.(2024)Yao, Su, and Yau]{doi:10.1073/pnas.2311436121}
Z.~Yao, J.~Su, and S.-T. Yau.
\newblock Manifold fitting with cyclegan.
\newblock \emph{Proceedings of the National Academy of Sciences}, 121\penalty0 (5):\penalty0 e2311436121, 2024.
\newblock \doi{10.1073/pnas.2311436121}.

\end{thebibliography}
}

\appendix
\newpage
\section{Proofs}
\subsection{Proof of Lemma 1}

\begin{proof}[Proof of Lemma 1]
    The marginal likelihood $p(y|k,\xi)$ is given by
    \begin{equation*}
        p(y|k,\xi)=p(y|Z) = \int_{(0,\infty)} \int_{\mathbb{R}^\nu} p(y|Z,\beta,\sigma) \pi(\beta|Z,\sigma)\pi(\sigma) d\beta d\sigma.
    \end{equation*}
    According to the specified priors, we have
    \begin{equation*}
        \begin{split}
            p(y|Z,\beta,\sigma) = \frac{1}{(2\pi\sigma^2)^{m/2}}\exp\{-\frac{1}{2\sigma^2}(y-Z\beta)^\top(y-Z\beta)\},\\
            \pi(\beta|Z,\sigma) = \frac{1}{(2\pi m \sigma^2)^{\nu/2}}|Z^\top Z|^{1/2}\exp \{ -\frac{1}{2m\sigma^2} \beta^\top Z^\top Z\beta\}.
        \end{split}
    \end{equation*}
    Then the Fubini Thm implies that,
    \begin{equation*}
        p(y|k,\xi) = \int_0^\infty \frac{1}{(2\pi \sigma^2)^{m/2} (m+1)^{\nu/2} } \exp \{-\frac{1}{2\sigma^2} a_{k,\xi}\} \pi(\sigma) d\sigma,
    \end{equation*}
    where $a_{k,\xi}=y^\top (I_m-\frac{m}{m+1}Z(Z^\top Z)^{-1}Z^\top)y$. With change of variables $w = \sigma / \sqrt{a_{k,\xi}}$, we have $p(y|k,\xi)\propto (m+1)^{-\nu/2}a_{k,\xi}^{-m/2}$. It follows from $\pi(k,\xi)\propto \tau(\mathcal{M}_k)^{-\gamma}$ that $p(k,\xi|y)\propto (m+1)^{-\nu/2}a_{k,\xi}^{-m/2}\tau(\mathcal{M}_k)^{-\gamma}$.
\end{proof}

\subsection{Proof of Lemma 3}

\begin{proof}[Proof of Lemma 3]
    According to (9),
    \begin{equation*}
        \alpha(k',\xi'|k,\xi) = \min \left\{1,~ \frac{p(k',\xi'|y)q(k,\xi|k',\xi')}{p(k,\xi|y)q(k',\xi'|k,\xi)}\right\}.
    \end{equation*}
    Notably, $\pi(k,\xi)q(k',\xi'|k,\xi)=\pi(k',\xi')q(k,\xi|k',\xi')$ under the priors and proposals. Then we have $\alpha(k',\xi'|k,\xi) = \min\{1,~ p(y|k',\xi')/p(y|k,\xi)\}$. Thus (6) implies this lemma. For the EBIC approximation, we substitute $\hat{p}$ for the corresponding $p$.
\end{proof}

\section{Additional simulation}

We conduct EBARS in the curve spline regression $(d=1,p=3)$ and the surface spline regression $(d=2,p=3)$. The performance is compared with BARS of \citet{10.1093/biomet/88.4.1055}, smoothing splines (SS) of \citet{green1994nonparametric} and thin plate splines (TPS) of \citet{10.1111/1467-9868.00374}. We calculate the predictive mean squared errors (MSE) for evaluation. Simulations show that the proposed method contributes to accurate predictions of all scenarios.

\subsection{Curve spline regression}

\begin{figure}
    \centering
    \includegraphics[width=0.85\textwidth]{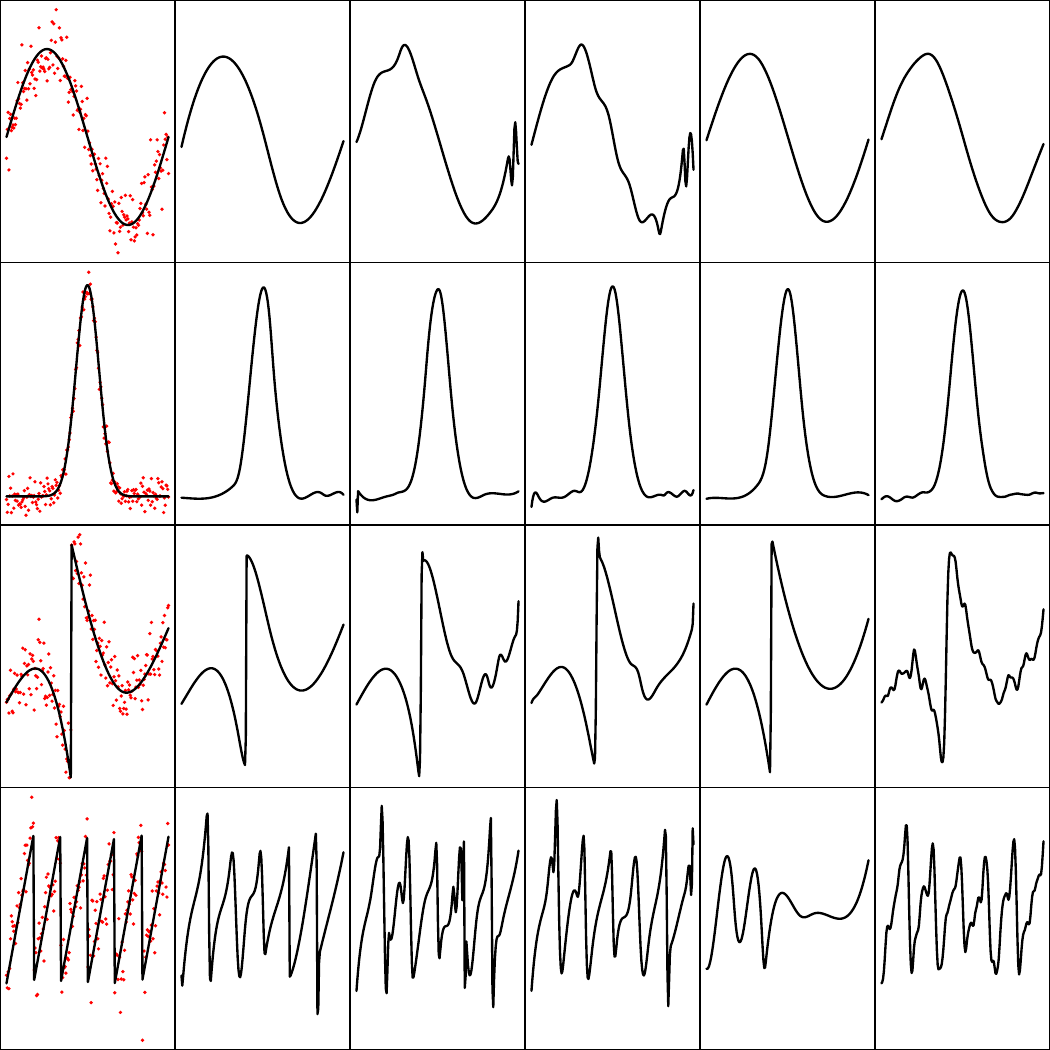}
    \caption{Curve spline regression. It corresponds to the prediction results with $m=200$ in Cases $1.1-1.4$. From left to right, columns are respectively the ground truth with red observed points, prediction by EBARS with $\gamma=1,0.5,0$, prediction by BARS and prediction by smoothing splines.}
    \label{fig:curve}
\end{figure}

%We conduct EBARS in curve spline regression and compare its prediction performance with BARS method \citep{10.1093/biomet/88.4.1055} and smoothing spline (SS) method \citep{wahba1990spline,green1994nonparametric}. 
The curves and data samples involved are illustrated in the first column of Figure \ref{fig:curve}. It can be seen that data are generated from $4$ different smoothness functions, denoted as Cases $1.1-1.4$ respectively. Cases $1.1$ and $1.2$ are continuous, whereas Cases $1.3$ and $1.4$ are discontinuous with one or multiple breakpoints. The outcome noise is Gaussian with standard deviation $2$, $2$, $4$, $1$. We compare the prediction performance with BARS and SS in curve fitting. To demonstrate the effect of $\gamma$ in EBIC, we implement $3$ versions of EBARS with $\gamma=1,0.5,0$. All methods are evaluated under sample sizes $m=200,500$ and the experiment is repeated $50$ times in each setting.

The mean squared errors are summarized in Table \ref{tab:curve}. To remove the effect of outliers, we drop out points with errors in the top or bottom $2.5\%$ to calculate censored MSE on test data. In EBARS, low $\gamma$ causes the model overfitting problem. It is clear that MSE rises up gradually as $\gamma$ decreases, especially in Case $1.3$. The behaviour is visualized in columns $2-4$ of Figure \ref{fig:curve}. This phenomenon is consistent with the theory. According to EBIC, the prior probability of the model space with $k$ knots satisfies $\pi(\mathcal{M}_k)\propto \tau(\mathcal{M}_k)^{1-\gamma}$. As $k<<n$ in practice, $\tau(\mathcal{M}_k)$ will grow with increasing number of knots. Consequently, the lower $\gamma$ means the higher $\pi(\mathcal{M}_k)$ for large $k$. Thereby more knots will be selected into the model, leading to overfitting in the end.

In this example, $\gamma=1$ gives the best performance among EBARS. Furthermore, BARS achieves similar results to EBARS $\gamma=1$ if the true knot number is relatively small, illustrated as in Cases $1.1-1.3$. However, when the required knot number is really large, BARS fails to work completely and attains a terrible MSE. See the $(4,5)$ panel of Figure \ref{fig:curve} for a graphical explanation. Besides, the smoothing spline method has a severe overfitting problem in Case $1.3$, as shown in the $(3,6)$ panel of Figure \ref{fig:curve}. This indicates that splines with many misplaced knots are inferior to adaptive knot splines in discontinuous cases.

\begin{table}
\caption{MSE on test data of five methods in Cases $1.1-1.4$}
\label{tab:curve}
\centering
\begin{tabular}{cccccccc}
\toprule
       Methods     &   $m$  & Case 1.1     & Case 1.2     & Case 1.3     & Case 1.4     \\ \midrule                               
                 
EBARS $\gamma=1$      & \multirow{5}{*}{$200$}     & 0.181(0.124) & 0.332(0.118) & 0.828(0.529) & 0.370(0.366)  \\
EBARS   $\gamma=0.5$  &       & 0.230(0.136)  & 0.389(0.184) & 1.412(0.783) & 0.368(0.150)  \\
EBARS   $\gamma=0$    &     & 0.341(0.229) & 0.464(0.201) & 1.960(0.985)  & 0.418(0.155) \\
BARS       &      & 0.178(0.150)  & 0.368(0.139) & 0.731(0.403) & 3.003(0.504) \\
SS &  & 0.137(0.088) & 0.329(0.132) & 3.772(1.687) & 0.443(0.155) \\
\midrule
EBARS $\gamma=1$     & \multirow{5}{*}{$500$}     & 0.068(0.033) & 0.147(0.061) & 0.361(0.168) & 0.079(0.033) \\
EBARS     $\gamma=0.5$   &    & 0.082(0.039) & 0.156(0.059) & 0.536(0.215) & 0.112(0.034) \\
EBARS    $\gamma=0$   &     & 0.170(0.073)  & 0.231(0.083) & 1.053(0.410)  & 0.144(0.042) \\
BARS        &     & 0.071(0.032) & 0.160(0.054)  & 0.317(0.140)  & 0.364(0.615) \\
SS & & 0.053(0.026) & 0.139(0.041) & 1.914(0.418) & 0.229(0.066) \\
\bottomrule
\end{tabular}
\end{table}
\subsection{Surface spline regression}

%We utilize tensor product splines to construct EBARS in bivariate spline regression and compare the performance with the BARS method and the thin plate spline (TPS) method of \citet{10.1111/1467-9868.00374}.
We compare the performance with BARS and TPS in the surface fitting.
%TPS is a popular spline-based approach for regression with multidimensional covariates.
The data is generated from four surfaces, denoted as Cases 2.1-2.4. Specifically, Cases 2.1 and 2.2 are smooth, whereas Cases 2.3 and 2.4 are discontinuous. Besides, Case 2.3 is with one jumping point in $x_1$ and Case 2.4 is with two jumping points in $x_1$ and $x_2$ separately. The outcome noise is Gaussian with standard deviation $1,1.5,2,2$. To remove outliers in discontinuous cases, we calculate MSE on test data without the upper or lower $2.5\%$ error. All methods are evaluated under sample sizes $m=500,1000$ and each experiment is repeated $20$ times.

\iffalse
\begin{figure}
    \centering
    \includegraphics[width=1\textwidth]{Experiments/plots/crop_surface_truth.pdf}
    \caption{The ground truth in surface spline regression for Cases $2.1-2.4$. Cases 2.1 and 2.2 are continuous, whereas Cases 2.3 and 2.4 are discontinuous.}
    \label{fig:surface_truth}
\end{figure}
\fi

The mean squared errors are summarized in Table \ref{tab:surface}. Three EBARS versions with $\gamma=1,0.5,0$ are implemented to demonstrate the effect of $\gamma$. In Cases 2.1 and 2.2, the performance of all five methods is satisfactory and MSE of TPS is roughly the smallest. This indicates that five spline approaches do well in fitting smooth surfaces. However, when the surface is discontinuous, the prediction performance of TPS is worse than free knot splines. For example, the MSE of TPS are much larger than that of EBARS and BARS in Cases 2.3 and 2.4. BARS achieves close results to EBARS except for Case 2.4. As $k$ has to be large enough to fit Case 2.4, the performance of BARS is limited owing to the poor property of BIC in high dimension. Besides, EBARS performs slightly better with low $\gamma$ than with high $\gamma$ even in smooth cases, different from curve regression. It seems that surface regression is more likely to be underfitting rather than overfitting.

\begin{table}
\caption{MSE on test data of five methods in Cases $2.1-2.4$}
\label{tab:surface}
\centering
\begin{tabular}{cccccc}
\toprule
Methods & m                     & Case 2.1     & Case 2.2     & Case 2.3     & Case 2.4     \\
\midrule
EBARS $\gamma=1$   & \multirow{5}{*}{500}  & 0.163(0.044) & 0.624(0.107) & 0.664(0.216) & 1.133(0.443) \\
EBARS $\gamma=0.5$  &                       & 0.148(0.038) & 0.616(0.111) & 0.597(0.192) & 1.101(0.287) \\
EBARS  $\gamma=0$ &                       & 0.138(0.042) & 0.609(0.089) & 0.568(0.136) & 1.076(0.269) \\
BARS    &                       & 0.169(0.050) & 0.671(0.160) & 0.660(0.199) & 1.968(0.485) \\
TPS     &                       & 0.102(0.031) & 0.587(0.092) & 1.958(0.625) & 2.324(0.466) \\
\midrule
EBARS $\gamma=1$  & \multirow{5}{*}{1000} & 0.064(0.011) & 0.355(0.051) & 0.230(0.061) & 0.331(0.077) \\
EBARS $\gamma=0.5$  &                       & 0.066(0.016) & 0.326(0.056) & 0.222(0.053) & 0.343(0.078) \\
EBARS $\gamma=0$  &                       & 0.065(0.014) & 0.306(0.058) & 0.242(0.067) & 0.336(0.058) \\
BARS    &                       & 0.107(0.027) & 0.416(0.094) & 0.313(0.074) & 1.020(0.510) \\
TPS     &                       & 0.057(0.015) & 0.319(0.032) & 1.205(0.215) & 1.379(0.214) \\
\bottomrule
\end{tabular}
\end{table}

\begin{figure}
    \centering
    \includegraphics[width=0.85\textwidth]{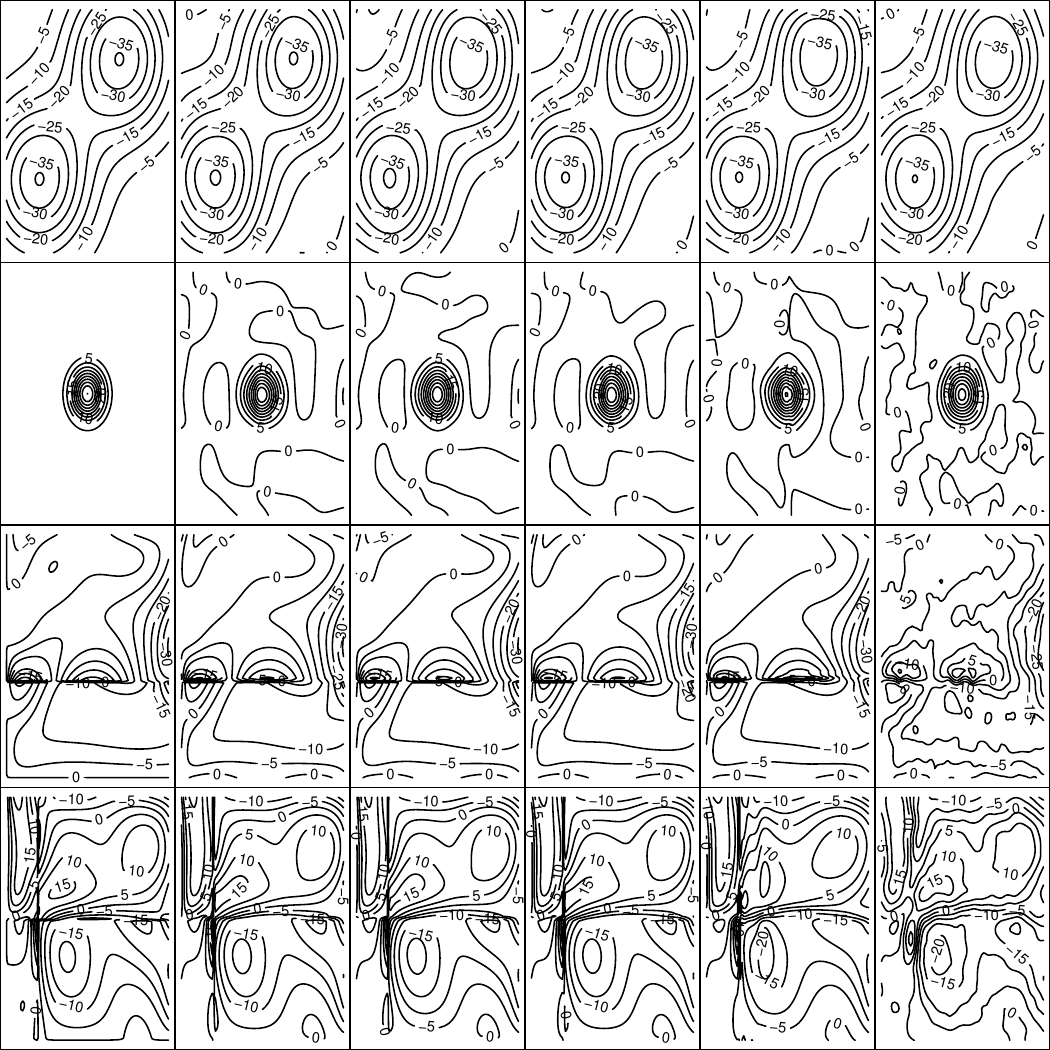}
    \caption{Surface spline regression. It illustrates the prediction results with $m=1000$ in Cases $2.1-2.4$ by contours. From left to right, columns are respectively the ground truth, prediction by EBARS with $\gamma=1,0.5,0$, prediction by BARS and prediction by TPS.}
    \label{fig:surface}
\end{figure}

To graphically illustrate the difference in surface spline regression, we visualize the prediction results with $m=1000$ by contour maps in Figure \ref{fig:surface}. Initially, the fitted contours approximately overlap with the ground truth in Cases 2.1 and 2.2, see the top $2$ rows. We can see from the $(3,6)$ and $(4,6)$ panels that TPS fails to estimate the jump discontinuity in Cases 2.3 and 2.4. Actually, the fitted surfaces by TPS are always smooth regardless of the smoothness of true models. BARS obtains a nice performance in Cases 2.1-2.3, but is trapped in Case 2.4 when the required $k$ is enormous. Conversely, Columns $2-4$  of Figure \ref{fig:surface} show that our EBARS makes precise predictions in all cases successfully.

\end{document}